\documentclass[showpacs,preprintnumbers,twocolumn,
amsmath,amssymb,groupedaddress,superscriptaddress]{revtex4}

\usepackage{graphicx}
\usepackage{dcolumn}
\usepackage{bm}

\begin{document}

\title{Temperature dependence of the volume and surface contributions to the nuclear symmetry energy
within the coherent density fluctuation model}

\author{A. N. Antonov}
\affiliation{Institute for Nuclear Research and Nuclear Energy,
Bulgarian Academy of Sciences, Sofia 1784, Bulgaria}

\author{D. N. Kadrev}
\affiliation{Institute for Nuclear Research and Nuclear Energy,
Bulgarian Academy of Sciences, Sofia 1784, Bulgaria}

\author{M. K. Gaidarov}
\affiliation{Institute for Nuclear Research and Nuclear Energy,
Bulgarian Academy of Sciences, Sofia 1784, Bulgaria}

\author{P. Sarriguren}
\affiliation{Instituto de Estructura de la Materia, IEM-CSIC,
Serrano 123, E-28006 Madrid, Spain}

\author{E. Moya de Guerra}
\affiliation{Grupo de F\'{i}sica Nuclear, Departamento de
F\'{i}sica At\'{o}mica, Molecular y Nuclear,\\ Facultad de
Ciencias F\'{i}sicas, Unidad Asociada UCM-CSIC(IEM), Universidad
Complutense de Madrid, E-28040 Madrid, Spain}

\begin{abstract}
The temperature dependence of the volume and surface components of the nuclear symmetry energy (NSE) and their ratio is investigated in the framework of the local
density approximation (LDA). The results of these quantities for finite nuclei are obtained within the coherent density fluctuation model (CDFM). The CDFM weight
function is obtained using the temperature-dependent proton and neutron densities calculated through the HFBTHO code that solves the nuclear
Skyrme-Hartree-Fock-Bogoliubov problem by using the cylindrical transformed deformed harmonic-oscillator basis. We present and discuss the values of the volume and
surface contributions to the NSE and their ratio obtained for the Ni, Sn, and Pb isotopic chains around double-magic $^{78}$Ni, $^{132}$Sn, and $^{208}$Pb nuclei.
The results for the $T$-dependence of the considered quantities are compared with estimations made previously for zero temperature showing the behavior of the NSE
components and their ratio, as well as with the available experimental data. The sensitivity of the results on various forms of the density dependence of the
symmetry energy is studied. We confirm the existence of ``kinks'' of these quantities as functions of the mass number at $T=0$ MeV for the double closed-shell
nuclei $^{78}$Ni and $^{132}$Sn and the lack of ``kinks'' for the Pb isotopes, as well as the disappearance of these kinks as the temperature increases.
\end{abstract}

\pacs{21.60.Jz, 21.65.Ef, 21.10.Gv}

\maketitle

\section{Introduction}
The study of the nuclear symmetry energy and particularly, its
density ($\rho$) and temperature ($T$) dependence is an important
task in nuclear physics. This quantity is related to the energy
connected with the conversion of the isospin asymmetric nuclear
matter (ANM) into a symmetric one. It is an important ingredient
of the nuclear equation of state (EOS) in a wide range of
densities and temperatures (see, e.g.,
\cite{NSE2014,Lattimer2007,Li2008}). Using approaches like the
local-density approximation
\cite{Agrawal2014,Samaddar2007,Samaddar2008,De2012} and the
coherent density fluctuation model
\cite{Gaidarov2011,Gaidarov2012,Gaidarov2014}, the knowledge of
EOS can give information about the properties of finite systems.
As noted in \cite{Bao2017} information about NSE from laboratory
experiments can be obtained from quantities that are sensitive to
it, such as static properties, nuclear excitations, collective
motions, heavy-ion reactions and others. One can add also
interesting phenomena including the supernova explosions,
properties of neutron stars and rare isotopes, frequencies and
strain amplitudes of gravitational waves from both isolated
pulsars and collisions involving neutron stars that depend
strongly on the EOS of neutron-rich nuclear matter. Here we would
like to mention a broad range of works devoted to the study of
$T$-dependence of single-particle properties of nuclear as well as
neutron matter (e.g.,
\cite{Agrawal2014,Moustakidis2007,Sammarruca2008,De2014,Zhang2014,Lee2010,BLi2006,Mekjian2005,Xu2007}),
and of $\rho$-dependence of NSE (e.g.,
\cite{Li2006,Piekarewicz2009,Vidana2009,Sammarruca2009,Dan2010}).
Among the important studies of the $T$-dependence of hot finite
nuclear systems we will note Ref.~\cite{Brack74} on thermal
Hartree-Fock (HF) calculations, the semiclassical approaches based
on the microscopic Skyrme-HF formalism \cite{Brack85}, the
Thomas-Fermi (TF) approximation \cite{Suraud87}, the inclusion of
the continuum effects in HF calculations at finite temperature
\cite{Bonche}, the refined TF approach \cite{De96} and extended TF
(ETF) method \cite{Brack84}. We mention also the
Hartree-Fock-Bogoliubov (HFB) models
\cite{Khan2007,Sandulescu2004,Monrozeau2007,Yuksel2014}, as well
as the relativistic TF approach with different relativistic
mean-field nuclear forces \cite{Zhang2014}. Here we should also
note the studies of the role of short-range and tensor
nucleon-nucleon correlations that change considerably the kinetic
and potential energy contributions to the NSE (e.g.
\cite{NSE2014,Bao2017}).

The method of the CDFM \cite{Ant80,AHP} allowed us to make the
transition from nuclear matter to finite nuclei in the studies of
the NSE for spherical \cite{Gaidarov2011} and deformed
\cite{Gaidarov2012} nuclei, as well as for Mg isotopes
\cite{Gaidarov2014} using the Brueckner energy-density
functional (EDF) of ANM \cite{Brueckner68,Brueckner69}.

In our previous work \cite{Antonov2017} we used a similar method to investigate the
$T$-dependence of the NSE for isotopic chains of even-even Ni, Sn,
and Pb nuclei following the LDA \cite{Agrawal2014,Samaddar2007,Samaddar2008,De2012}
and using instead of the Brueckner EDF,
the Skyrme EDF with SkM* and SLy4 forces.
The $T$-dependent local densities $\rho(r,T)$ and kinetic energy
densities $\tau(r,T)$ were calculated within a self-consistent
Skyrme HFB method using the cylindrical transformed deformed
harmonic-oscillator basis (HFBTHO)
\cite{Stoitsov2013,Stoitsov2005} with the same forces. For comparison, the kinetic
energy density $\tau$ was calculated also by the TF expression
(for low temperatures) up to $T^{2}$ term \cite{Lee2010}. In
addition, we studied the $T$-dependence of NSE for $^{208}$Pb
using the ETF method \cite{Brack85,Brack84} and the rigorous
density functional approach \cite{Stoitsov87} to obtain $T$-dependent local density distributions.

In his pioneering work \cite{Feenberg47} Feenberg in 1947 showed
that the surface energy should contain a symmetry energy
contribution due to the failure of nuclear saturation at the edge
of the nucleus and that the volume saturation energy also has a
symmetry energy term. In later works
\cite{Cameron57,Bethe71,Green58,Myers66} were given analyses of
the volume and surface components of the NSE, as well as of their
ratio. We should also mention some works in which the volume and
surface components to NSE and their ratio are explicitly related to
the neutron-skin thickness
\cite{Dan2003,Dan2004,Dan2014,Dan2009,Dan2011,Tsang2012,Tsang2009,Ono2004,Dan2006,Warda2010,Centelles2010}.
Agrawal {\it et al.} \cite{Agrawal2014} pointed out the
substantial change in the NSE coefficients for finite nuclei with
temperature in comparison with the case of nuclear matter. Lee and
Mekjian \cite{Lee2010} emphasized the bigger sensitivity of the
surface component of NSE to the temperature in respect to the
volume energy term. We note also studies of the volume and surface
NSE, e.g., in
Refs.~\cite{Steiner2005,Danielewicz,Diep2007,Kolomietz,Nikolov2011,Jiang2012,Ma2012},
including those by analyzing the experimental nuclear binding
energies \cite{Jiang2012,Ma2012}.

The values of the temperature-dependent symmetry energy
coefficients for sixty nine spherical and non-spherical nuclei
with mass $36\leq A \leq 218$ and charge $14\leq Z \leq 92$
numbers, respectively, have been evaluated in Ref.~\cite{De2012a}
in the subtracted finite-temperature TF framework by using two
different energy-density functionals. A substantial temperature
dependence of the surface symmetry energy was found, while the
volume symmetry energy turned out to be less sensitive to the
temperature. As a result, the symmetry energy coefficient of
finite nuclei falls down as the temperature rises \cite{De2012a},
an observation that has been confirmed in Ref.~\cite{Antonov2017}.
A study of the decomposition of NSE into spin and isospin
components is carried out in Ref.(\cite{Guo2018}) within the
Brueckner-Hartree-Fock approximation with two- and three-body
forces.

In our work \cite{Antonov2016} the volume and surface
contributions to the NSE and their ratio were calculated within
the CDFM using two EDF's, namely the Brueckner
\cite{Brueckner68,Brueckner69} and Skyrme (see
Ref.~\cite{Wang2015}) ones. The CDFM weight function was obtained
by means of the proton and neutron densities obtained from the
self-consistent deformed HF+BCS method with density-dependent
Skyrme interactions. The obtained results in the cases of Ni, Sn,
and Pb isotopic chains were compared with results of other
theoretical methods and with those from other approaches which
used experimental data on binding energies, excitation energies to
isobaric analog states (IAS), neutron-skin thicknesses and with
results of other theoretical methods. We note that in
\cite{Antonov2016} the obtained values of the volume and surface
components of NSE and their ratio concern the case at $T=0$ MeV.

The aim of the present work is to evaluate the above mentioned quantities for temperatures different from zero. The $T$-dependent local density distributions
$\rho_{p}(r,T)$ and $\rho_{n}(r,T)$ computed by the HFBTHO code are used to calculate the $T$-dependent CDFM weight function. Such an investigation of the thermal
evolution of the NSE components and their ratio for isotopes belonging to the Ni, Sn, and Pb chains around the double-magic nuclei, will extend our previous
analysis of these nuclei considering them as cold systems \cite{Antonov2016}. At the same time, the obtained results within the CDFM provide additional information
on the thermal mapping of the volume and surface symmetry energies that has been poorly investigated till now (e.g., Ref.~\cite{De2012a}). In addition, we study the
sensitivity of the calculated $T$-dependent quantities on different available forms of the density dependence of the symmetry energy.

The structure of this paper is the following. In
Sec.~\ref{sec:formalism} we present the main relationships for the
NSE and its volume and surface components depending on the
temperature that we use in our study, as well as the CDFM
formalism that provides a way to calculate the mentioned
quantities. Section~\ref{sec:results} contains the numerical
results and discussions. The main conclusions of the study are
given in Sec.~\ref{sec:conclusions}.

\section{Theoretical formalism}
\label{sec:formalism}

We start this section with the expression for the nuclear energy
given in the droplet model that is an extension of the
Bethe-Weizs\"{a}cker liquid drop model to incorporate the surface
asymmetry. It can be written as \cite{Steiner2005,Myers1969}:
\begin{eqnarray}
E(A,Z)=&-&BA+E_SA^{2/3}+S^VA\frac{(1-2Z/A)^2}{1+S^SA^{-1/3}/S^V} \nonumber \\
&+& \displaystyle E_C \frac{Z^2}{A^{1/3}} +E_{dif} \frac{Z^2}{A} +
E_{ex}\frac{Z^{4/3}}{A^{1/3}} \nonumber\\
&+& a\Delta A^{-1/2}
\label{eq:01}
\end{eqnarray}
In Eq.~(\ref{eq:01}) $B \simeq 16$ MeV is the binding energy per
particle of bulk symmetric matter at saturation. $ E_S $, $ E_C $,
$ E_{dif} $, and $ E_{ex} $ are coefficients that correspond to
the surface energy of symmetric matter, the Coulomb energy of a
uniformly charged sphere, the diffuseness correction and the
exchange correction to the Coulomb energy, while the last term
gives the pairing corrections ($\Delta$ is a constant and $a=+1$
for odd-odd nuclei, 0 for odd-even and -1 for even-even nuclei).
$S^V$ is the volume symmetry energy parameter and $S^S$ is the
modified surface symmetry energy one in the liquid model (see
Ref.~\cite{Steiner2005}, where it is defined by $S^{S*}$).

In our previous work \cite{Antonov2017} we studied the temperature
dependence of the NSE $S(T)$. For the aims of the present work we
will rewrite the symmetry energy (the third term in the right-hand
side of Eq.~(\ref{eq:01}) in the form
\begin{equation}
S(T)\frac{(N-Z)^2}{A}
\label{eq:02}
\end{equation}
where
\begin{equation}
S(T)=\frac{S^V(T)}{1 + \displaystyle \frac{S^S(T)}{S^V(T)}A^{-1/3}}=\frac{S^V(T)}{1+A^{-1/3}/\kappa(T)}
\label{eq:03}
\end{equation}
with
\begin{equation}
\kappa (T) \equiv \frac{S^V(T)}{S^S(T)}.
\label{eq:04}
\end{equation}
In the case of nuclear matter, where $A \longrightarrow \infty$
and $S^S/S^V \longrightarrow 0$, we have $S(T)=S^V(T) $. Also at
large $A$ Eq.~(\ref{eq:03}) can be written in the known form (see
Ref.~\cite{Bethe71}):
\begin{equation}
S(T)=\frac{S^V(T)}{1 + \displaystyle \frac{S^S(T)}{S^V(T)}A^{-1/3}} \simeq c_{3}-\frac{c_{4}}{A^{1/3}},
\label{eq:05}
\end{equation}
where $ c_3=S^V $ and $ c_4=S^S $. From Eq.~(\ref{eq:03}) follow
the relations of $S^V(T)$ and $S^S(T)$ with $S(T)$:
\begin{equation}
S^V(T)=S(T)\left(1+\frac{1}{\kappa(T) A^{1/3}}\right),
\label{eq:06}
\end{equation}
\begin{equation}
S^S(T)=\frac{S(T)}{\kappa(T)}\left(1+\frac{1}{\kappa (T) A^{1/3}}\right) .
\label{eq:07}
\end{equation}

In what follows we use essentially the CDFM scheme to calculate
the NSE and its components (see
Refs.~\cite{Ant80,AHP,Antonov2016}) in which the one-body density
matrix $\rho(\mathbf{r},\mathbf{r}^{\prime})$ is a coherent superposition of the
one-body density matrices $\rho_{x}({\bf r},{\bf r^{\prime}})$ for
spherical "pieces" of nuclear matter ("fluctons") with densities
$\rho_{x}({\bf r})=\rho_{0}(x)\Theta(x-|{\bf r}|)$ and
$\rho_{0}(x)=3A/4\pi x^{3}$. It has the form:
\begin{equation}
\rho({\bf r},{\bf r^{\prime}})=\int_{0}^{\infty}dx |{\cal
F}(x)|^{2} \rho_{x}({\bf r},{\bf r^{\prime}})
\label{eq:08}
\end{equation}
with
\begin{eqnarray}
\rho_{x}({\bf r},{\bf r^{\prime}})&=&3\rho_{0}(x)
\frac{j_{1}(k_{F}(x)|{\bf r}-{\bf r^{\prime}}|)}{(k_{F}(x)|{\bf
r}-{\bf r^{\prime}}|)}\nonumber \\  & \times & \Theta \left
(x-\frac{|{\bf r}+{\bf r^{\prime}}|}{2}\right ),
\label{eq:09}
\end{eqnarray}
where $j_{1}$ is the first-order spherical Bessel function and
\begin{equation}
k_{F}(x)=\left(\frac{3\pi^{2}}{2}\rho_{0}(x)\right )^{1/3}\equiv
\frac{\beta}{x}
\label{eq:10}
\end{equation}
with
\begin{equation}
\beta=\left(\frac{9\pi A}{8}\right )^{1/3}\simeq 1.52A^{1/3}
\label{eq:11}
\end{equation}
is the Fermi momentum of the nucleons in the flucton with a radius
$x$. The density distribution in the CDFM has the form:
\begin{equation}
\rho({\bf r})=\int_{0}^{\infty}dx|{\cal
F}(x)|^{2}\rho_{0}(x)\Theta(x-|{\bf r}|).
\label{eq:12}
\end{equation}
It follows from (\ref{eq:12}) that in the case of monotonically
decreasing local density ($d\rho/dr\leq 0$) the weight function
$|{\cal F}(x)|^{2}$ can be obtained from a known density
(theoretically or experimentally obtained):
\begin{equation}
|{\cal F}(x)|^{2}=-\frac{1}{\rho_{0}(x)} \left.
\frac{d\rho(r)}{dr}\right |_{r=x}.
\label{eq:13}
\end{equation}

We have shown in our previous works
\cite{Gaidarov2011,Gaidarov2012,Antonov2016} that the NSE in the
CDFM for temperature $T=0$ can be obtained in the form:
\begin{equation}
S=\int_{0}^{\infty}dx|{\cal F}(x)|^{2}S[\rho(x)],
\label{eq:14}
\end{equation}
where the symmetry energy for the asymmetric nuclear matter that
depends on the density $S[\rho(x)]$ has to be determined using a
chosen EDF (in \cite{Antonov2016} Brueckner and Skyrme EDF's have
been used).

In this work the $T$-dependent NSE $S(T)$ is calculated by the
expressions similar to Eq.~(\ref{eq:14}) but containing
$T$-dependent quantities:
\begin{equation}
S(T)=\int_{0}^{\infty}dx|{\cal F}(x,T)|^{2}S[\rho(x,T)].
\label{eq:15}
\end{equation}
In Eq.~(\ref{eq:15}) the weight function $|{\cal F}(x,T)|^{2}$
depends on the temperature through the temperature-dependent total
density distribution $\rho_{total}(r,T)$:
\begin{equation}
|{\cal F}(x,T)|^{2}=-\frac{1}{\rho_{0}(x)} \left.
\frac{d\rho_{total}(r,T)}{dr}\right |_{r=x},
\label{eq:16}
\end{equation}
where
\begin{equation}
\rho_{total}(r,T)=\rho_{p}(r,T)+\rho_{n}(r,T),
\label{eq:17}
\end{equation}
$\rho_{p}(r,T)$ and $\rho_{n}(r,T)$ being the proton and neutron
$T$-dependent densities that in our work \cite{Antonov2017} were
calculated using the HFB method with transformed
harmonic-oscillator basis and the HFBTHO code \cite{Stoitsov2013}.

As mentioned, in the present work we consider the $T$-dependence
of the NSE $S(T)$, but also of its volume $S^V(T)$ and surface
$S^S(T)$ components and their ratio $\kappa(T)$
[Eq.~(\ref{eq:04})].

Following Refs.~\cite{Dan2003,Dan2006,Diep2007,Antonov2016} an
approximate expression for the ratio $\kappa(T)$ can be written
within the CDFM:
\begin{equation}
\kappa(T) =\frac{3}{R\rho_{0}}\int_{0}^{\infty}dx |{\cal F}(x,T)|^{2} x \rho_{0}(x)
\left\{\frac{S(\rho_{0})}{S[\rho(x,T)]}-1\right \},
\label{eq:18}
\end{equation}
where $|{\cal F}(x,T)|^{2}$ is determined by Eq.~(\ref{eq:16}), $R=r_0A^{1/3}$ \cite{Diep2007} and $S(\rho_0)$ is the NSE at equilibrium nuclear matter density
$\rho_0$ and $T=0$ MeV. For instance, the values of $S(\rho_0)$ for different Skyrme forces in the Skyrme EDF are given in Table II of Ref.~\cite{Antonov2016}. In
what follows we use the commonly employed power parametrization (firstly, in subsection \ref{subsec:our}) for the density dependence of the symmetry energy (e.g.,
\cite{Diep2007,Dan2004,Dan2006})
\begin{equation}
S[\rho(x,T)]= S^V(T) \left[ \frac{\rho(x,T)}{\rho_0}
\right]^{\gamma}.
\label{eq:19}
\end{equation}
There exist various estimations for the value of the parameter
$\gamma$. For instance, in Ref.~\cite{Diep2007} $\gamma=0.5 \pm
0.1$ and in Ref.~\cite{Dan2006} $0.54 \leq \gamma \leq 0.77$. The
estimations in Ref.~\cite{samm17} (given in Table 2 there) of the
NSE based on different cases within the chiral effective field
theory and from other predictions are $\gamma=0.60 \pm 0.05$
(N$^2$LO), $\gamma=0.55 \pm 0.03$ (N$^3$LO), $\gamma=0.55$ (DBHF)
and 0.79 (APR \cite{akmal98}). Another estimation of $\gamma=0.72
\pm 0.19$ is also given in Ref.~\cite{russ16}.

Using Eq.~(\ref{eq:19}) (and having in mind that $S(\rho_0)=S^V$),
Eqs.~(\ref{eq:15}) and (\ref{eq:18}) can be re-written as follows:
\begin{equation}
S(T)=S(\rho_0) \int_{0}^{\infty}dx|{\cal F}(x,T)|^{2} \left[
\frac{\rho(x,T)}{\rho_0} \right]^{\gamma},
\label{eq:20}
\end{equation}
\begin{eqnarray}
\kappa (T) \equiv \frac{S^V(T)}{S^S(T)}& = &\frac{3}{R\rho_{0}}\int_{0}^{\infty}dx |{\cal F}(x,T)|^{2} x \rho_{0}(x) \nonumber\\
&\times& \left\{\left[ \frac{\rho_{0}}{\rho(x,T)}\right]^{\gamma}-1\right \}. \label{eq:21}
\end{eqnarray}

\section{Results of calculations and discussion}
\label{sec:results}
\subsection{Results with the density dependence of the symmetry energy given by Eq.~(\ref{eq:19})}
\label{subsec:our}

The calculations of the $T$-dependent nuclear symmetry energy
$S(T)$, its volume $S^V(T)$ and surface $S^S(T)$ components, as
well as their ratio $\kappa(T) = S^V(T)/S^S(T)$ were performed
using the relationships (\ref{eq:15}), (\ref{eq:18})-(\ref{eq:21})
with the weight function $|{\cal F}(x,T)|^2$ from
Eqs.~(\ref{eq:16}) and (\ref{eq:17}). As mentioned in
Sec.~\ref{sec:formalism}, the $T$-dependent proton $\rho_p(r,T)$
and neutron $\rho_n(r,T)$ density distributions, as well as the
total density $\rho_{total}(r,T)$  [Eq.~(\ref{eq:17})] were
calculated using the HFBTHO code from Ref.~\cite{Stoitsov2013}
with the Skyrme EDF for SkM* and SLy4 forces. We note that in the
calculations of $S(T)$ [Eq.~(\ref{eq:20})] and $\kappa(T)$
[Eq.~(\ref{eq:21})] we use the weight function $|{\cal F}(x,T)|^2$
from Eq.~(\ref{eq:16}), where the density distributions for finite
nuclei are used. The quantity $S[\rho(x,T)]$ in Eqs.~(\ref{eq:15})
and (\ref{eq:18}) is the symmetry energy for asymmetric nuclear
matter chosen in the parametrized form of Eq.~(\ref{eq:19}). Our
calculations are performed for the Ni, Sn, and Pb isotopic chains.

Studying the $T$-dependence of the mentioned quantities we
observed (as can be seen below) a certain sensitivity of the
results to the value of the parameter $\gamma$ in
Eq.~(\ref{eq:19}). In order to make a choice of its value we
imposed the following physical conditions: i) the obtained results
for the considered quantities at $T=0$ MeV to be equal or close to
those obtained for the same quantities in our previous works for
the NSE, its components and their ratio
(Ref.~\cite{Antonov2017,Antonov2016}), and ii) their  values for
$T=0$ MeV to be compatible with the available experimental data
(see, e.g., the corresponding references in  \cite{Antonov2016}).

\begin{figure*}
\centering
\includegraphics[width=0.75\linewidth]{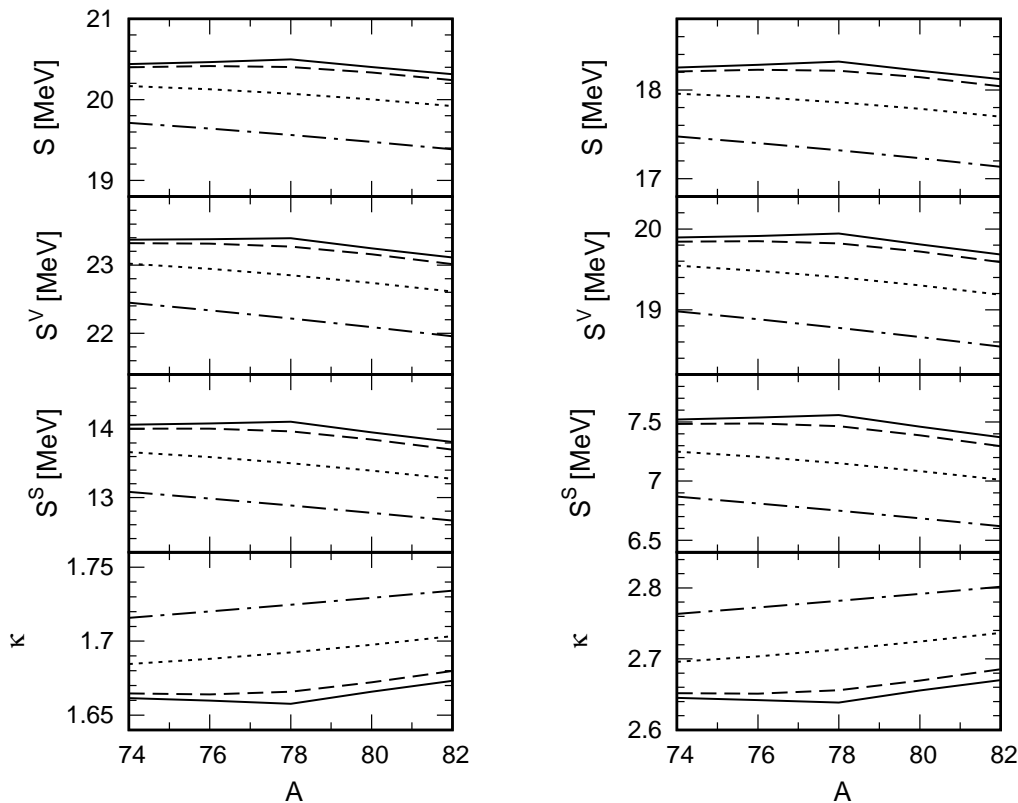}
\caption[]{Mass dependence of the NSE $S(T)$, its volume $S^V(T)$
and surface $S^S(T)$ components and their ratio $\kappa(T)$ for
nuclei from the Ni isotopic chain at temperatures $T=0$ MeV (solid
line), $T=1$ MeV (dashed line), $T=2$ MeV (dotted line), and $T=3$
MeV (dash-dotted line) calculated with SkM* Skyrme interaction for
values of the parameter $\gamma=0.3$ (left panel) and $\gamma=0.4$
(right panel). \label{fig1}}
\end{figure*}
\begin{figure*}
\centering
\includegraphics[width=0.75\linewidth]{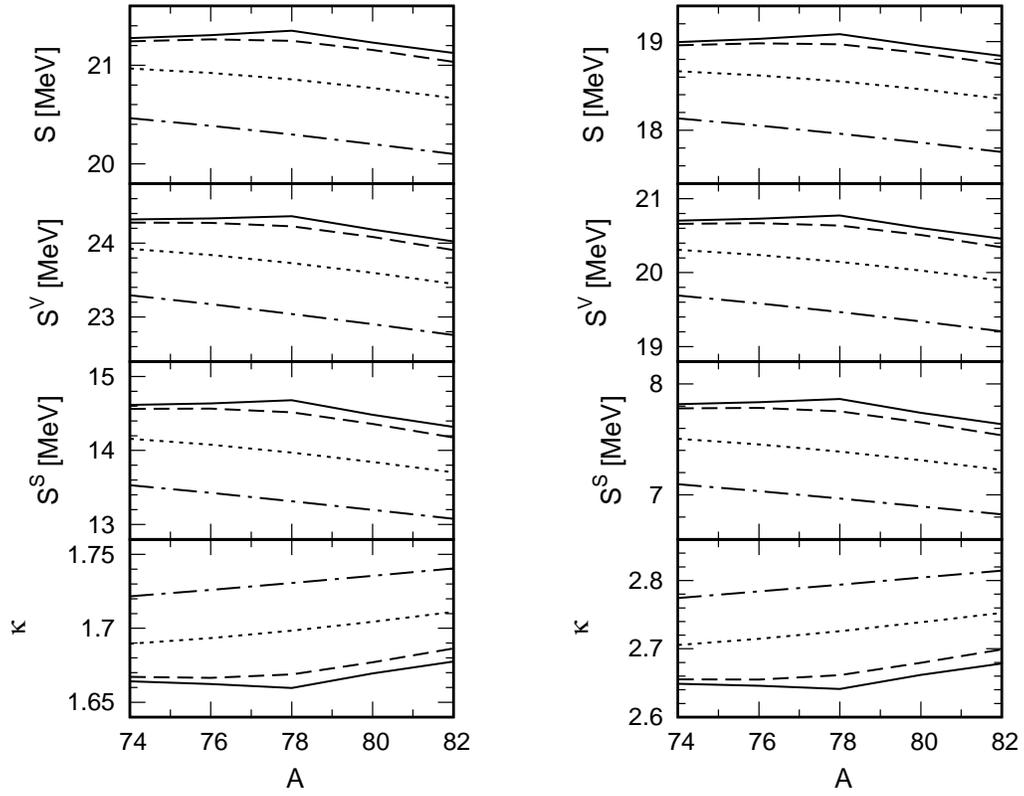}
\caption[]{Same as in Fig.~\ref{fig1}, but with SLy4 Skyrme interaction. \label{fig2}}
\end{figure*}
\begin{figure*}
\centering
\includegraphics[width=0.75\linewidth]{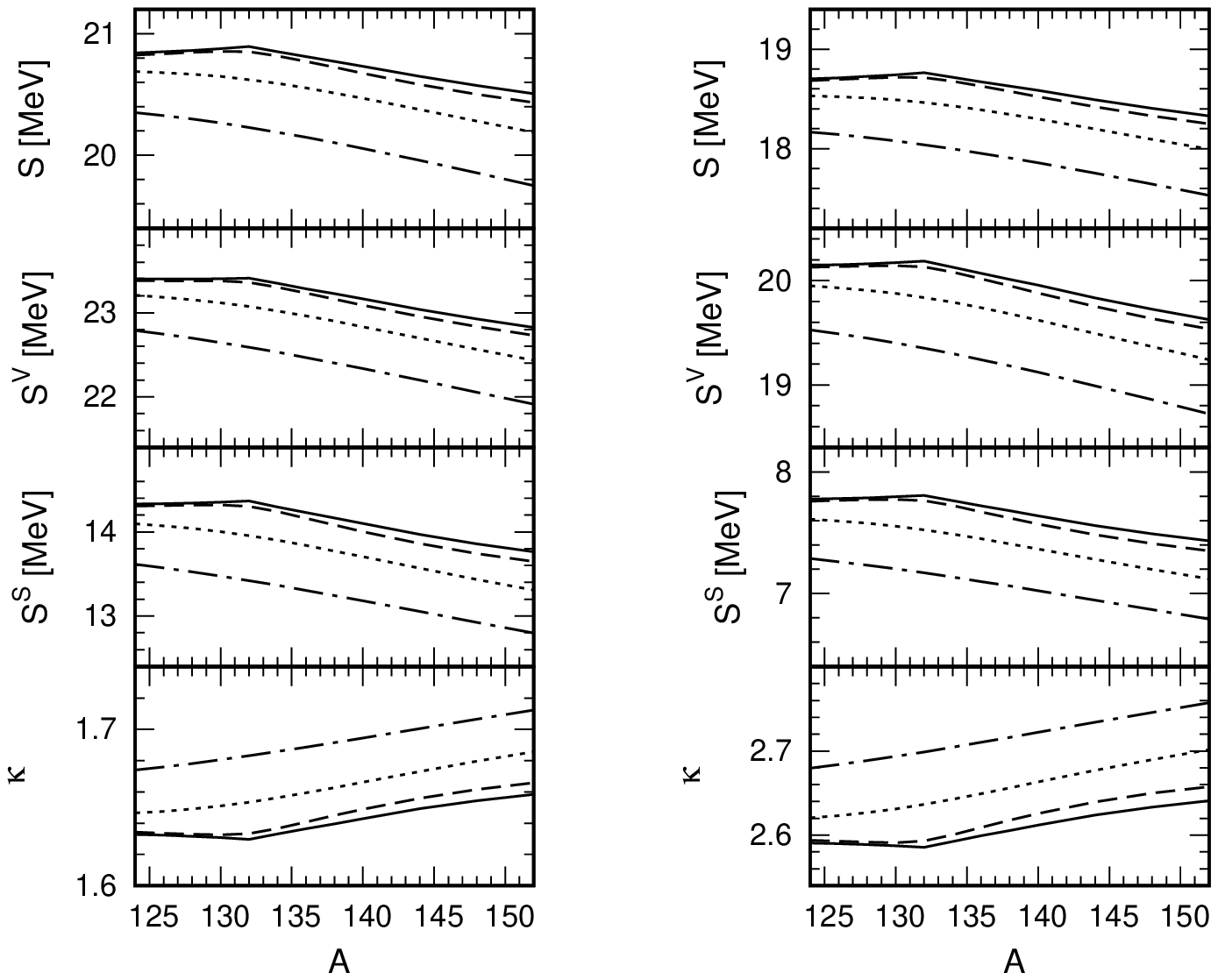}
\caption[]{Same as in Fig.~\ref{fig1}, but for nuclei from the Sn isotopic chain. \label{fig3}}
\end{figure*}
\begin{figure*}
\centering
\includegraphics[width=0.75\linewidth]{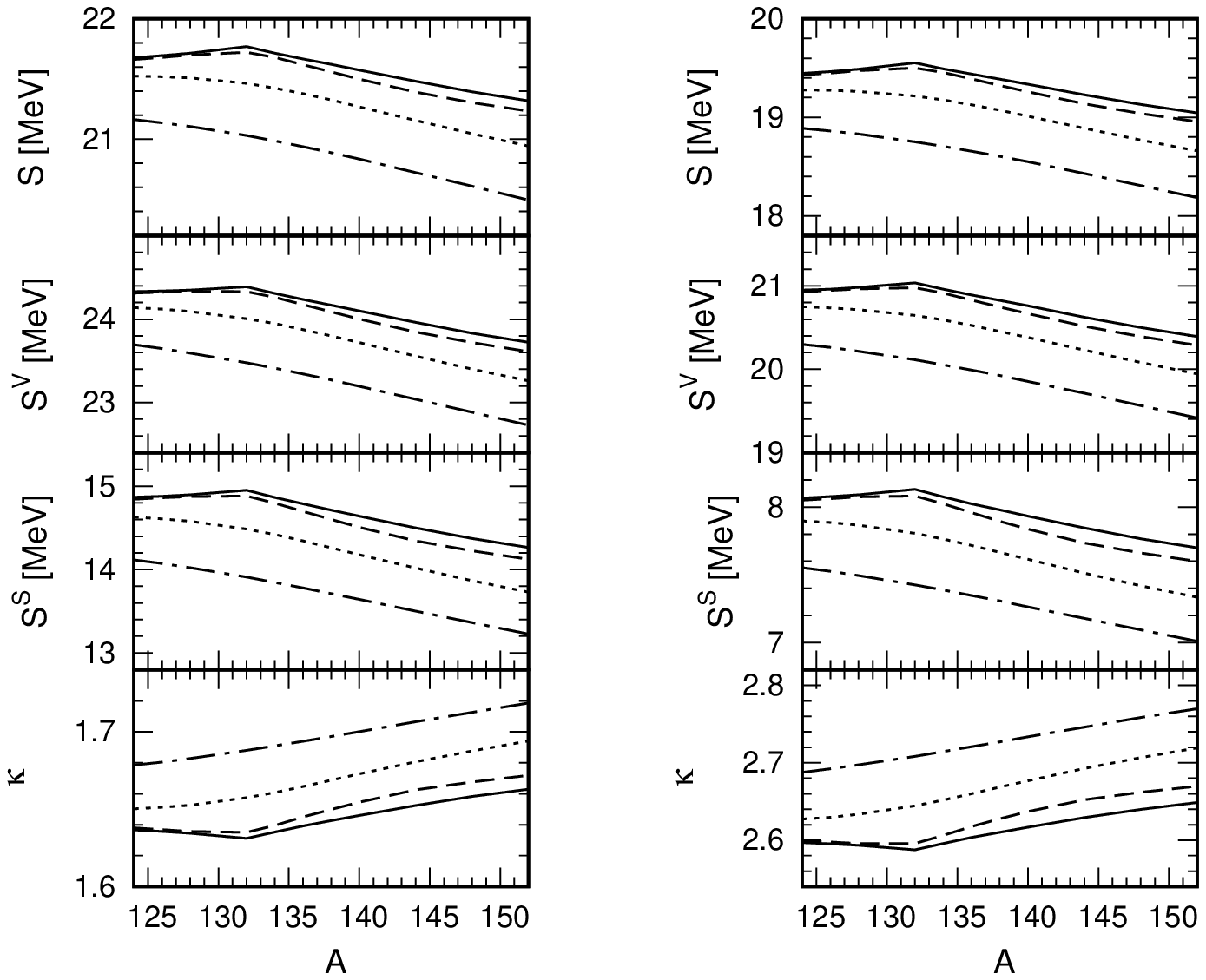}
\caption[]{Same as in Fig.~\ref{fig2}, but for nuclei from the Sn isotopic chain. \label{fig4}}
\end{figure*}
\begin{figure*}
\centering
\includegraphics[width=0.75\linewidth]{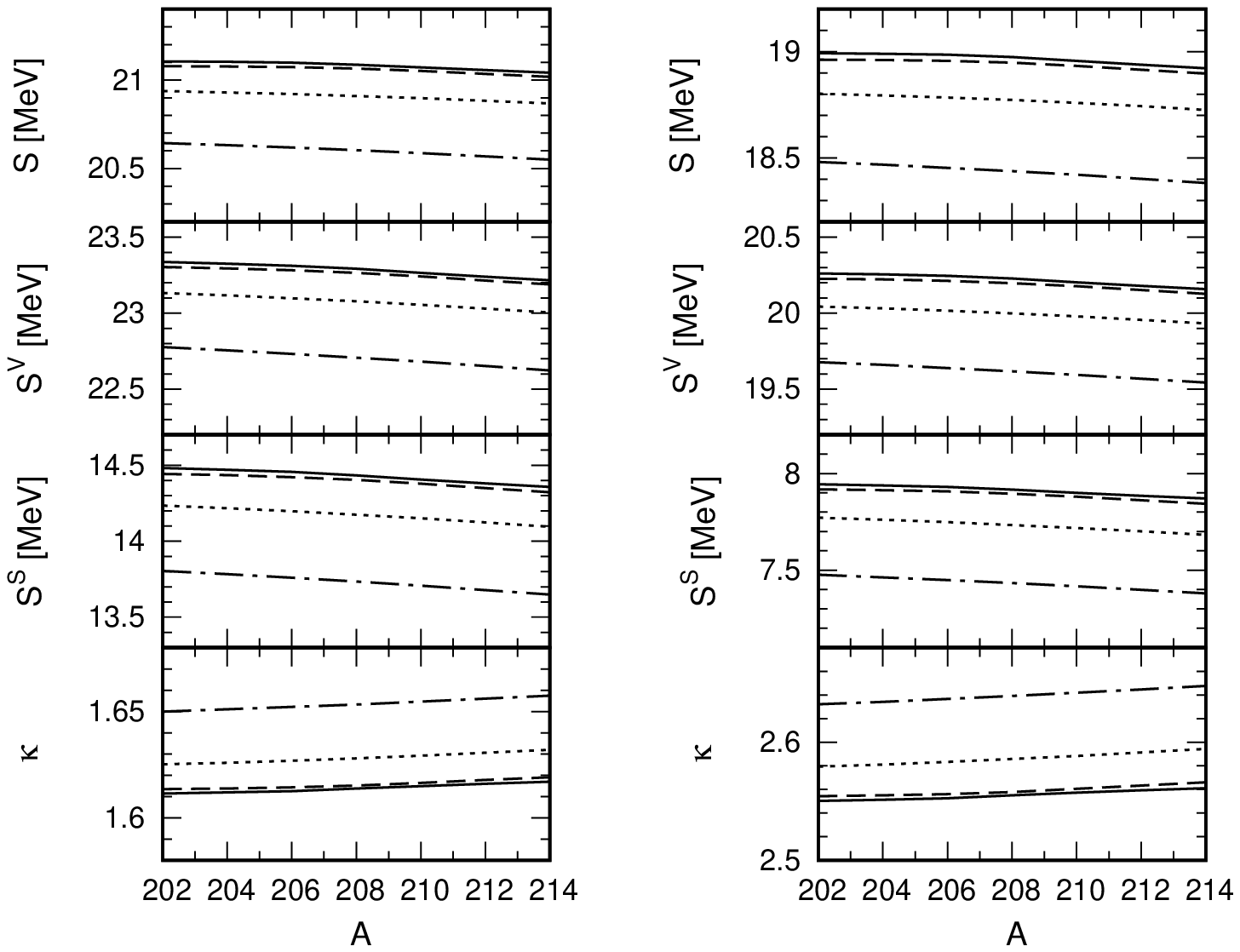}
\caption[]{Same as in Fig.~\ref{fig1}, but for nuclei from the Pb isotopic chain. \label{fig5}}
\end{figure*}
\begin{figure*}
\centering
\includegraphics[width=0.75\linewidth]{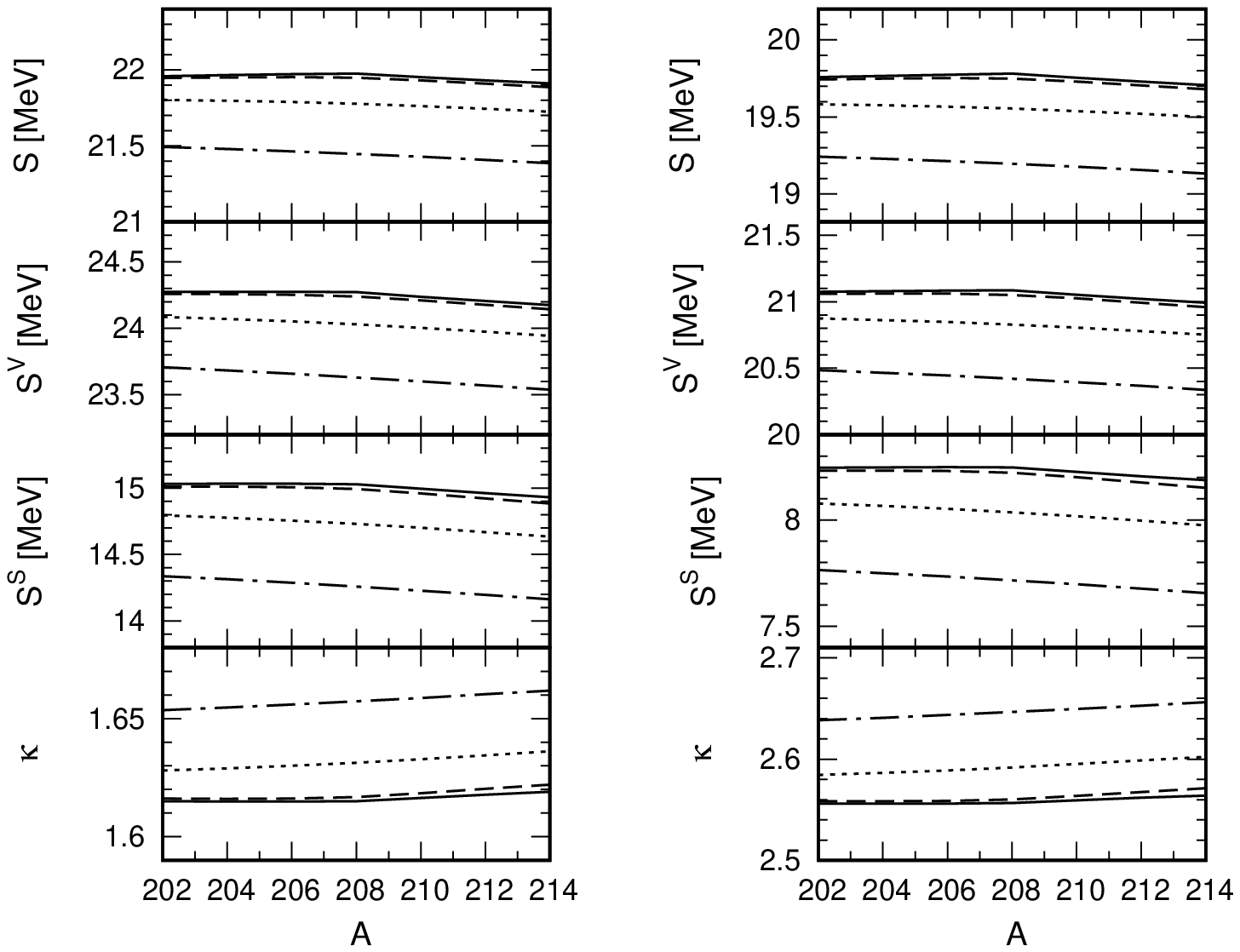}
\caption[]{Same as in Fig.~\ref{fig2}, but for nuclei from the Pb isotopic chain. \label{fig6}}
\end{figure*}

In Figs.~\ref{fig1}-\ref{fig6} the results for $ S(T) $, $ S^V(T)
$, $ S^S(T) $, and $\kappa = S^V(T)/S^S(T)$ are given as functions
of the mass number $A$ for the isotopic chains of Ni, Sn, and Pb
nuclei for temperatures $T=0$--3 MeV calculated using the SkM* and
SLy4 Skyrme forces. The results are presented for two values of
the parameter $\gamma=0.3$ and $0.4$. The reason for this choice
is related to the physical criterion mentioned above. It can be
seen that at $T=0$ MeV and $\gamma=0.4$ the value of $\kappa$ is
around $2.6$. This result is in agreement with our previous result
obtained in the case of the Brueckner EDF in
Ref.~\cite{Antonov2016}, namely $2.10\le \kappa\le 2.90$. The
latter is compatible with the published values of $ \kappa $
extracted from nuclear properties presented in
Ref.~\cite{Diep2007} from the IAS and skins \cite{Dan2004} ($2.6
\le \kappa\le 3.0$) and from masses and skins \cite{Dan2003}
($2.0\le \kappa\le  2.8$). In the case of $\gamma =0.3$ our result
for $T=0$ MeV is $\kappa = 1.65$ that is in agreement with the
analyses of data in Ref.~\cite{Diep2007} ($1.6\le \kappa\le 2.0$),
as well as with the results of our work \cite{Antonov2016} in the
case of Skyrme EDF, namely, for the Ni isotopic chain $ 1.5 \le
\kappa\le 1.7 $, for the Sn isotopic chain $ 1.52\le \kappa\le 2.1
$, and for the Pb isotopic chain $ 1.65\le \kappa\le  1.75 $, all
obtained with SLy4 and SGII forces.

Before making the comparison of our results for $S(T)$ at $T=0$
MeV with our previous results from Ref.~\cite{Antonov2017} (there
the NSE is denoted by $ e_{sym} $) we mention that though the
latter are in good agreement with theoretical  predictions for
some specific nuclei reported by other  authors, we showed that
they depend on the suggested  definitions of this quantity. The
comparison of the results in the present work for $S$ at $T=0$ MeV
with those from our work Ref.~\cite{Antonov2017} (the latter
illustrated there in Fig.~12) shows that they agree with our
present values of $S$ within the range of $\gamma=0.3$--0.4,
except in the case of Pb with SLy4 force, where the present
results are somewhat lower.

It can be seen from Figs.~\ref{fig1}--\ref{fig6} that the
quantities $S(T)$, $S^V(T)$, and $S^S(T)$ decrease with increasing
temperatures ($T=0$--3 MeV), while $\kappa(T)$ slowly increases
when $T$ increases. This is true for both Skyrme forces (SkM* and
SLy4) and for the three isotopic chains of the Ni, Sn, and Pb
nuclei. Here we would like to note that the values of $ \gamma $
between 0.3 and 0.4 that give an agreement of the studied
quantities with data, as well as with our previous results for
$T=0$ MeV, are in the lower part of the estimated limits of the
values of $\gamma$ (e.g., in the case of $\gamma = 0.5\pm 0.1$
\cite{Diep2007}).

It can be seen also from Figs.~\ref{fig1}--\ref{fig6} that there
are ``kinks'' in the curves of $S(T)$, $S^V(T)$, $S^S(T)$, and
$\kappa(T)$ for $T=0$ MeV in the cases of double closed-shell
nuclei $^{78}$Ni and $^{132}$Sn and no ``kinks'' in the Pb chain.
This had been observed also in our previous work for $S(T)$
\cite{Antonov2017}, as well as for its volume and surface
components and their ratio $\kappa$ at $T=0$ MeV in
Ref.~\cite{Antonov2016}.

\begin{figure*}
\centering
\includegraphics[width=0.75\linewidth]{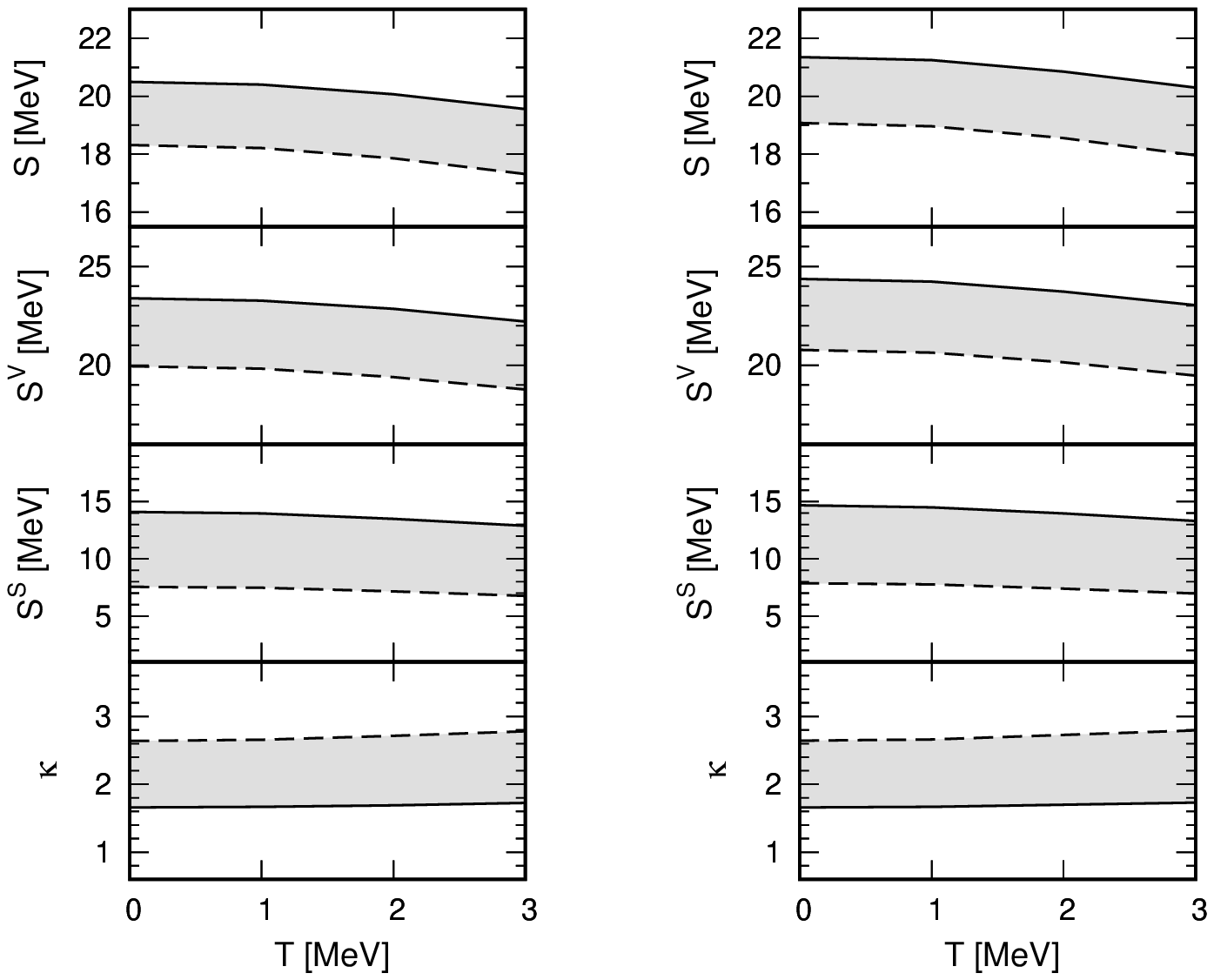}
\caption[]{Temperature dependence of the NSE $S(T)$, its volume
$S^V(T)$ and surface $S^S(T)$ components, and their ratio
$\kappa(T)$ obtained for values of the parameter $\gamma=0.3$
(solid line) and $\gamma=0.4$ (dashed line) with SkM* (left panel)
and SLy4 (right panel) forces for $^{78}$Ni nucleus. \label{fig7}}
\end{figure*}
\begin{figure*}
\centering
\includegraphics[width=0.75\linewidth]{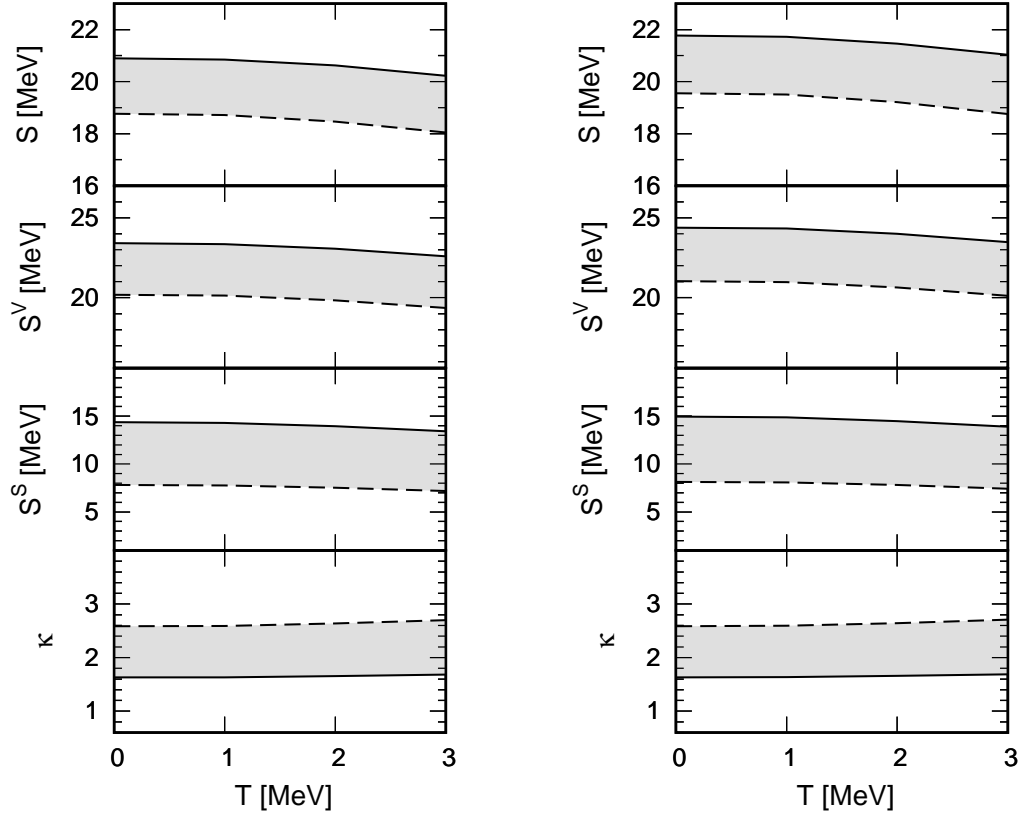}
\caption[]{Same as in Fig.~\ref{fig7}, but for $^{132}$Sn nucleus. \label{fig8}}
\end{figure*}
\begin{figure*}
\centering
\includegraphics[width=0.75\linewidth]{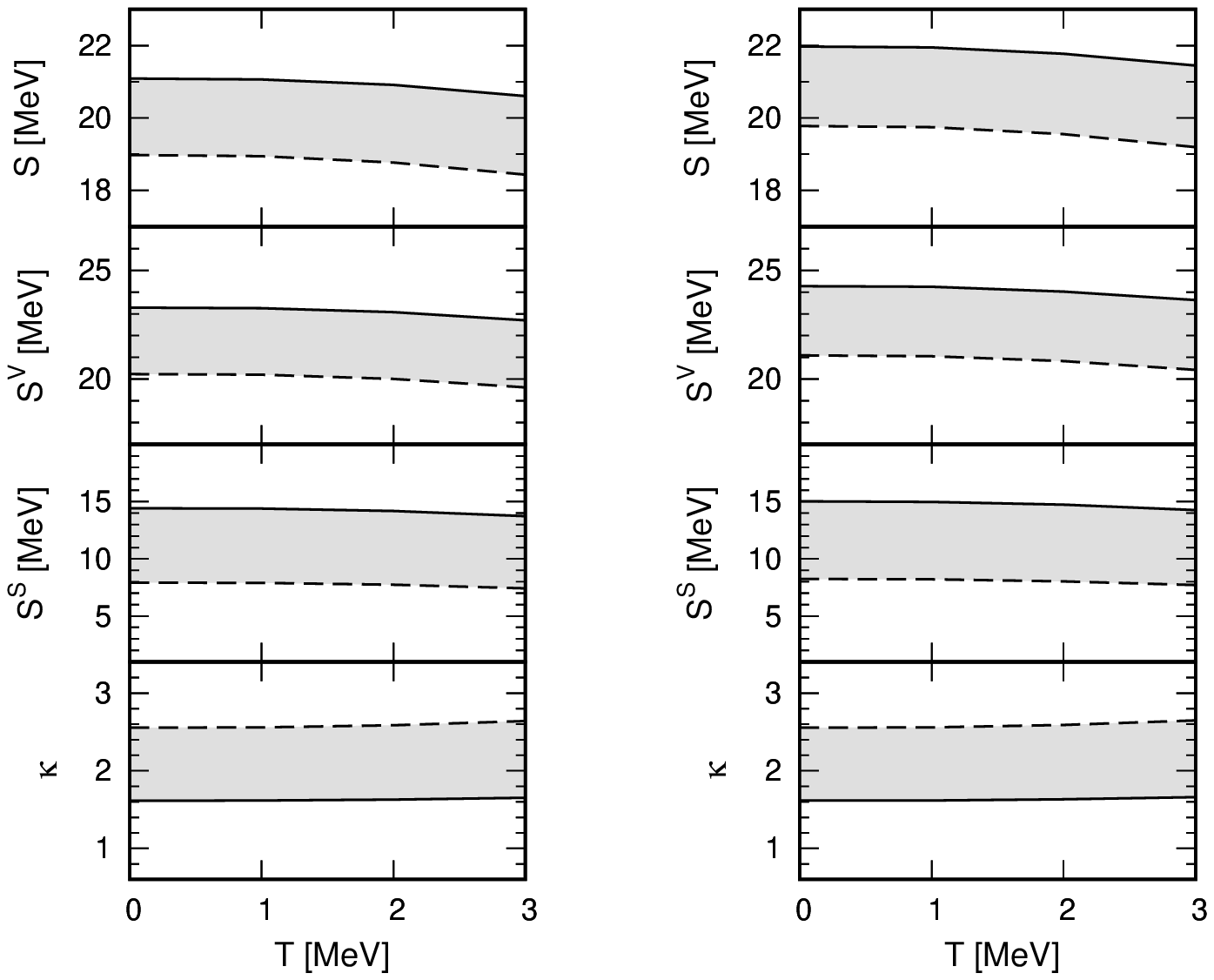}
\caption[]{Same as in Fig.~\ref{fig7}, but for $^{208}$Pb nucleus. \label{fig9}}
\end{figure*}

In Figs.~\ref{fig7}--\ref{fig9} are given the results for the
$T$-dependence of $S(T)$, $S^V(T)$, $S^S(T)$, and $\kappa(T)$ for
the double-magic $^{78}$Ni, $^{132}$Sn, and $^{208}$Pb nuclei
obtained using both SkM* and SLy4 Skyrme forces. The results are
presented by grey areas between the curves for the values of the
parameter $\gamma = 0.3$ and $\gamma= 0.4$. It can be seen that
$S(T)$, $S^V(T)$, and $S^S(T)$ decrease, while $\kappa(T)$ slowly
increases with the increase of the temperature for both Skyrme
forces.

\subsection{Comparison with results from alternative parametrizations of
the density dependence of the symmetry energy}
\label{subsec:comparison}

In this part of our work we study the sensitivity of the obtained results towards different density dependences of the symmetry energy. First, we note the Eq.~(6)
from Ref.~\cite{Dong2012}:
\begin{equation}
S(\rho)=C_{k}\left(\frac{\rho}{\rho_{0}}\right)^{2/3}+ C_{1}\left(\frac{\rho}{\rho_{0}}\right)+ C_{2}\left(\frac{\rho}{\rho_{0}}\right)^{1.52}.
\label{eq:22}
\end{equation}
This relationship, which coincides with the shape from the density-dependent M3Y interaction \cite{Muk2007}, is similar to that of Eq.~(9) of
Ref.~\cite{Agrawal2013}, where the last term of Eq.~(\ref{eq:22}) is exchanged by $C_{2}(\rho/\rho_{0})^{5/3}$. Second, we use also the dependence of the symmetry
energy \cite{Tsang2009,Dong2012}
\begin{equation}
S(\rho)=12.5\left(\frac{\rho}{\rho_{0}}\right)^{2/3}+17.6\left(\frac{\rho}{\rho_{0}}\right)^{\gamma}.
\label{eq:23}
\end{equation}
It was noted in \cite{Tsang2009} that the analysis of both isospin diffusion and double ratio data involving neutron and proton spectra by an improved quantum
molecular dynamics transport model suggests values of $\gamma=0.4-1.05$.

Concerning the values of the coefficients $C_{1}$ and $C_{2}$ in Eq.~(\ref{eq:22}), we determined them by fitting the three curves of $S(\rho)$ (for different
values of $S(\rho_{0})$) presented in Fig.~4 of Ref.~\cite{Dong2012} and note them as $S_{1}(\rho)$ when  $S(\rho_{0})=29.4$ MeV, $S_{2}(\rho)$ when
$S(\rho_{0})=31.6$ MeV, and $S_{3}(\rho)$ when  $S(\rho_{0})=33.8$ MeV.

We mention also the remark in Ref.~\cite{Dong2012} that the expressions for $S(\rho)$ in Eqs.~(\ref{eq:19}) and (\ref{eq:23}) "do not reproduce the density
dependence of the symmetry energy as predicted by the mean-field approach around nuclear saturation density". Nevertheless, in what follows we present for a
comparison our results from subsection \ref{subsec:our} with those calculated not only using Eq.~(\ref{eq:22}) but also $S(\rho)$ from Eq.~(\ref{eq:23}) which we
will note as $S_{4}(\rho)$.

\begin{figure}
\centering
\includegraphics[width=80mm]{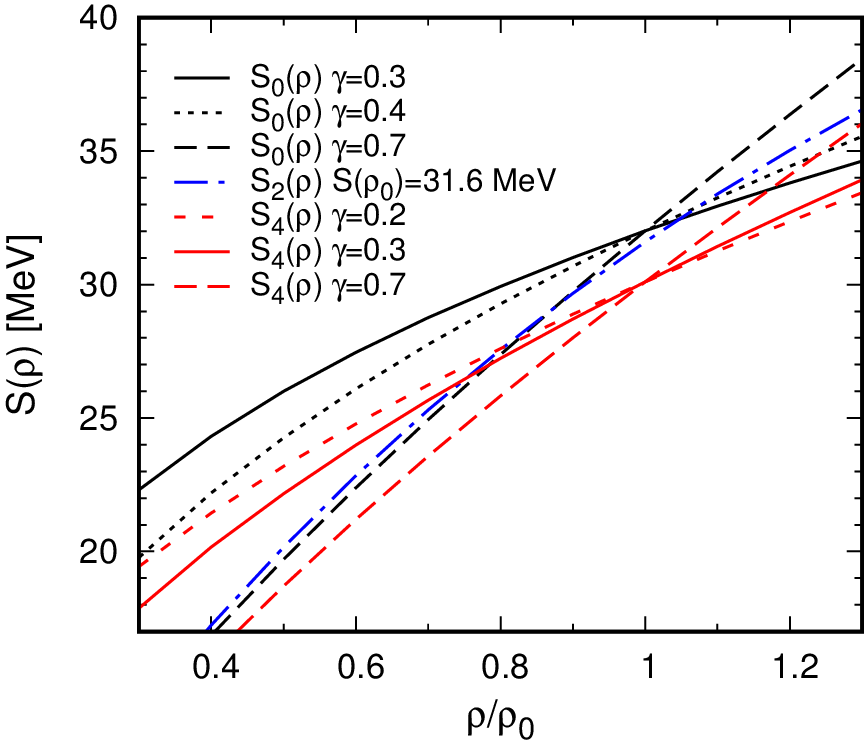}
\caption[]{(Color online) Behavior of the density-dependent symmetry energy: $S_{2}(\rho)$ with $S(\rho_{0})=31.6$ MeV [Eq.~(\ref{eq:22}) with $C_{k}=17.47$,
$C_{1}=27.94$, and $C_{2}=-13.81$] (blue dash-dotted line); $S_{4}(\rho)$ [Eq.~(\ref{eq:23})] with $\gamma=0.2$ (red short-dashed line), $\gamma=0.3$ (red solid
line), $\gamma=0.7$ (red dashed line), and $S_{0}(\rho)$ [Eq.~(\ref{eq:19})] with $\gamma=0.3$ (black solid line), $\gamma=0.4$ (black dotted line), $\gamma=0.7$
(black dashed line). \label{fig10}}
\end{figure}

Figure~\ref{fig10} illustrates the behavior of the density dependence of the symmetry energy $S(\rho)$ by giving some of the curves for different functions, namely,
$S_{2}(\rho)$, three curves using Eq.~(\ref{eq:19}) that we label as $S_{0}(\rho)$ with $\gamma=0.3$, 0.4, 0.7, as well as three curves for $S_{4}(\rho)$ that
correspond to $\gamma=0.2$, 0.3, 0.7. At this point we emphasize that the CDFM weight function $|{\cal F}(x,T)|^2$ which is used in Eqs.~(\ref{eq:15}) and
(\ref{eq:18}) has a form of a bell with a maximum around $x=R_{1/2}$ at which the value of the density $\rho(x=R_{1/2})$ is half of the value of the central density
equal to $\rho_{0}$ [$\rho(R_{1/2})/\rho_{0}=0.5$]. So, namely in this region (around $\rho/\rho_{0}=0.5$), the values of the different $S(\rho)$ play the main role
in the calculations of $S(T)$ [Eq.~(\ref{eq:15})] and $\kappa(T)$ [Eq.~(\ref{eq:18})].

In the next Fig.~\ref{fig11} we consider, as an example, the mass dependence of $ S(T) $, $ S^V(T) $, $ S^S(T) $, and $\kappa = S^V(T)/S^S(T)$ in the case of the Ni
isotopic chain for temperatures $T=0$--3 MeV using the SLy4 Skyrme force. The results are given when the symmetry energy has the form of $S_{4}(\rho)$ at
$\gamma=0.2$ and 0.3. One can see that, e.g., the value of $\kappa$ at $T=0$ MeV and $\gamma=0.2$ is 1.90 that is close to the result obtained using
Eqs.~(\ref{eq:19}) and (\ref{eq:21}) for $\gamma=0.3$ (it is $\kappa=1.66$) shown in the left panel of Fig.~\ref{fig2}. The value of $\kappa$ when $\gamma=0.3$ is
2.69 which is similar to that in the case of Eq.~(\ref{eq:19}) and (\ref{eq:21}) for $\gamma=0.4$ (it is $\kappa=2.64$) shown in the right panel of Fig.~\ref{fig2}.
There exist in these cases similarities also of the behavior of the quantities $ S(T) $, $ S^V(T) $, and $ S^S(T) $ as functions of $T$ in the corresponding cases.
The reason for the mentioned similarities is the closeness of the corresponding curves shown in Fig.~\ref{fig10} in the region around $\rho/\rho_{0}=0.5$.

\begin{figure*}
\centering
\includegraphics[width=0.75\linewidth]{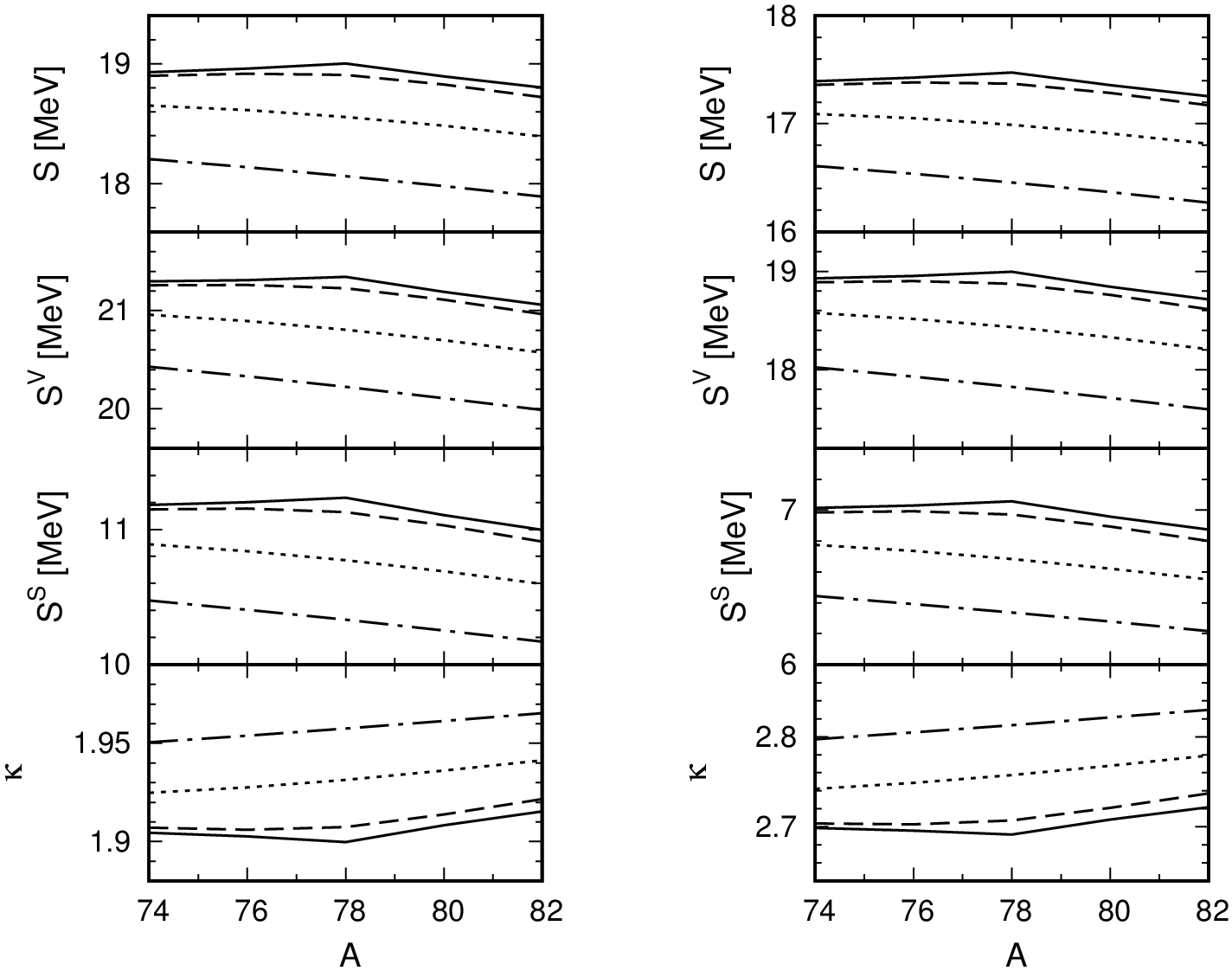}
\caption[]{Same as in Fig.~\ref{fig2}, but with use of $S_{4}(\rho)$ [Eq.~(\ref{eq:23})] for values of the parameter $\gamma=0.2$ (left panel) and $\gamma=0.3$
(right panel). \label{fig11}}
\end{figure*}

\begin{figure}
\centering
\includegraphics[width=0.75\linewidth]{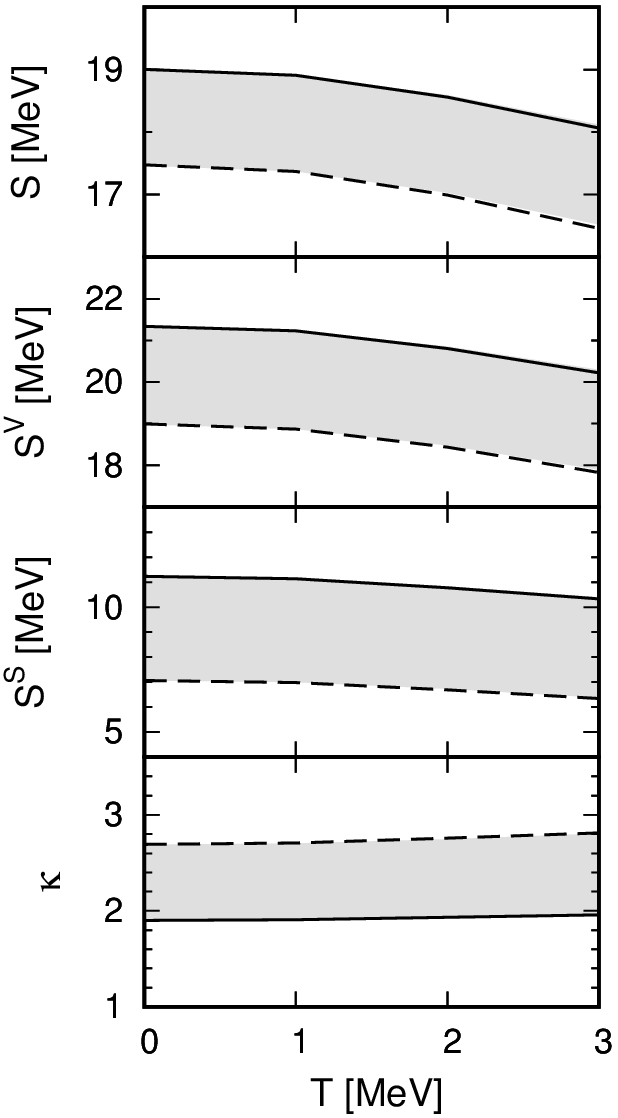}
\caption[]{Same as in Fig.~\ref{fig7} (right panel), but with use of $S_{4}(\rho)$ [Eq.~(\ref{eq:23})] for values of the parameter $\gamma=0.2$ (solid line) and
$\gamma=0.3$ (dashed line). \label{fig12}}
\end{figure}

In Fig.~\ref{fig12} we give, as an example, the $T$-dependence of $ S(T) $, $ S^V(T) $, $ S^S(T) $, and $\kappa$ in the case of $^{78}$Ni nucleus using the density
dependence of the symmetry energy  $S_{4}(\rho)$ for $\gamma=0.2$ and $\gamma=0.3$ in the case of SLy4 force. This figure corresponds to Fig.~\ref{fig7} (right
panel) and one can see the similarities of the presented quantities. As we mentioned before, the reason for the latter is the closeness of the corresponding curves
at around $\rho/\rho_{0}=0.5$.

Table~\ref{tab1} reflects the noted above similarities of the results for $\kappa$ and $S(T)$ at $T=0$ MeV in the cases when Eqs.~(\ref{eq:19}) and (\ref{eq:23})
are used. However, the results obtained for $\kappa$ using the density-dependence from Eq.~(\ref{eq:22}) are quite different from those in Table~\ref{tab1} with
either set of parameter values: $\kappa=10.2$ for $S_{1}(\rho)$, $\kappa=11.9$ for $S_{2}(\rho)$, and $\kappa=13.82$ for $S_{3}(\rho)$ (while the values of $S(T)$
for the same symmetry energy forms are 14.62 MeV, 14.22 MeV, and 13.85 MeV, respectively). The reason for this comes from the fact that the integrand in
Eq.~(\ref{eq:18}) for $\kappa$ is strongly peaked around $\rho(R_{1/2})/\rho_{0}=0.5$, and the results depend on the behavior of $S_{2}(\rho)$ in respect to
$S_{0}(\rho)$ and $S_{4}(\rho)$ in this region (see Fig.~\ref{fig10}). It is also worth mentioning that the large values of the ratio $\kappa$ given above are
comparable with those presented in Table~\ref{tab1} at large values of the parameter $\gamma$ ($\gamma=0.7$).

\begin{table}
\caption{Values of the parameters $\gamma$, the ratio $\kappa$ and the symmetry energy $S(T)$ (in MeV) for $^{78}$Ni nucleus at $T=0$ MeV in the case of the density
dependence of the symmetry energy given by Eq.~(\ref{eq:19}) (left part) and by $S_{4}(\rho)$ [Eq.~(\ref{eq:23})] (right part).} \label{tab1}
\begin{center}
\begin{tabular}{ccccccccccccccc}
\hline \hline \noalign{\smallskip}
$\gamma$ & & $\kappa$ & & $S(T)$ & & & & & & $\gamma$ & & $\kappa$ & & $S(T)$  \\
\noalign{\smallskip}\hline\noalign{\smallskip}
0.3 & & 1.66  & & 21.35  & & & & & & 0.2 & & 1.90 & &  19.00 \\
0.4 & & 2.64  & & 19.08  & & & & & & 0.3 & & 2.69 & &  17.48 \\
0.5 & & 4.09  & & 17.20  & & & & & & 0.4 & & 3.74 & &  16.23 \\
0.7 & & 10.67 & & 14.26  & & & & & & 0.7 & & 9.82 & &  13.58 \\
\noalign{\smallskip}\hline \hline
\end{tabular}
\end{center}
\end{table}

\section{Conclusions}
\label{sec:conclusions}

In the present work we perform calculations of the temperature
dependence of the NSE $S(T)$, its volume $S^V(T)$ and surface
$S^S(T)$ components, as well as their ratio $\kappa(T)=
S^V(T)/S^S(T)$. Our method is based on the local density
approximation. It uses the coherent density fluctuation model
\cite{Ant80,AHP} with $T$-dependent proton $\rho_{p}(r,T)$,
neutron $\rho_{n}(r,T)$, and total
$\rho_{total}(r,T)=\rho_{p}(r,T)+\rho_{n}(r,T)$ density
distributions. The latter are calculated within the
self-consistent Skyrme HFB method using the cylindrical
transformed harmonic-oscillator basis (HFBTHO)
\cite{Stoitsov2013,Stoitsov2005} and the corresponding code with
 SkM* and SLy4 Skyrme forces. In the calculations we used in Eqs.~(\ref{eq:15}) and (\ref{eq:18}) different density dependences of the symmetry energy
 $S[\rho(x,T)]$, namely Eq.~(\ref{eq:19}) in subsection \ref{subsec:our} and the alternative cases, Eqs.~(\ref{eq:22}) and (\ref{eq:23}) in subsection
 \ref{subsec:comparison}. The quantities of interest are
 calculated for the isotopic chains of Ni, Sn, and Pb nuclei.

The main results of the present work can be summarized as follows:

(i) With increasing $T$, the quantities $S$, $S^V$, and $S^S$ decrease, while $\kappa$ slightly increases for all the isotopes in the three chains for both Skyrme
forces and for all used density dependences of the symmetry energy. The same conclusion can be drawn for the thermal evolution of the mentioned quantities in the
cases of the three considered double-magic $^{78}$Ni, $^{132}$Sn, and $^{208}$Pb nuclei.

(ii) The results for $S(T)$, $S^V(T)$, $S^S(T)$, and $\kappa(T)$ are sensitive to the choice of the density dependence of the symmetry energy $S[\rho(x,T)]$ in
Eqs.~(\ref{eq:15}) and (\ref{eq:18}). In subsection \ref{subsec:our} the sensitivity of the studied quantities towards the values of the parameter $\gamma$ in
Eq.~(\ref{eq:19}) is shown. In subsection \ref{subsec:comparison} are considered in detail the results when other different density dependences of the symmetry
energy [Eqs.~(\ref{eq:19}), (\ref{eq:22}), and (\ref{eq:23})] are used. The similarities and differences between the results from various functional forms are
related to the behavior of the corresponding values of $S[\rho(x,T)]$ around the value of the ratio $\rho(x,T)/\rho_{0}=0.5$ for which the CDFM weight function has
a maximum.

(iii) In the cases of double-magic $^{78}$Ni and $^{132}$Sn nuclei we observe ``kinks'' for $T=0$ MeV in the curves of $S(T)$, $S^V(T)$, $S^S(T)$, and $\kappa(T)$,
but not in the case of Pb isotopes. This effect was also observed in our previous works. It is also worth mentioning how the kinks are blurred and eventually
disappear as $T$ increases, demonstrating its close relationship with the shell structure.

\begin{acknowledgments}
Three of the authors (M.K.G., A.N.A., and D.N.K) are grateful for support of
the Bulgarian Science Fund under Contract No.~DFNI-T02/19. E.M.G. and P.S. acknowledge
support from MINECO (Spain) under Contract FIS2014--51971--P(E.M.G. and P.S.) and
FPA2015-65035-P(E.M.G.).
\end{acknowledgments}

\end{document}